\documentclass[pra,aps,twocolumn,showpacs,10pt,floatfix]{revtex4-1}
\usepackage{amsmath,amssymb,graphicx,bm,color,ulem}
\usepackage{amsfonts,amsthm}

\begin{document}

\title{Broadband X-waves with orbital angular momentum}

\author{Miguel A. Porras}
\affiliation{Grupo de Sistemas Complejos, ETSIME, Universidad Polit\'ecnica de Madrid, Rios Rosas 21, 28003 Madrid, Spain}

\author{Raúl García-Álvarez}

\affiliation{Grupo de Sistemas Complejos, ETSIME, Universidad Polit\'ecnica de Madrid, Rios Rosas 21, 28003 Madrid, Spain}

\begin{abstract}
We describe the minimal diffraction-free and dispersion-free, superluminal wave packets (X-waves and generic superluminal localized waves) that can carry a given amount of units $m$ of orbital angular momentum (OAM) per photon. Even if their frequency spectrum is wide enough to synthesize a unipolar pulse, OAM imposes a temporal pulse shape with as many zeroes as units of OAM, and therefore $|m|/2$ temporal oscillations. All the frequencies in the broadband spectrum are displayed, like in a rainbow, from the bluer ones in the vicinity of the vortex to the redder ones in the wave periphery. At the same time, the whole wave paclet experiences a blue shift proportional to $|m|$. As a result of the radial red shift and the OAM blue shift, the colour of the brilliant ring around the central vortex is independent of OAM, and solely determined by the broadband source spectrum. Given the very peculiar properties of X-waves with OAM, their generation and use, instead of standard Laguerre-Gauss modes, could improve the performance of OAM-based communication and quantum cryptographic systems, as well as the efficiency of the generation of high harmonics and attosecond pulses with OAM.
\end{abstract}


\maketitle



\section{Introduction}

X-waves are diffraction-free and dispersion-free wavepackets localized in space and time travelling at arbitrary superluminal group velocities in free space \cite{LU,SAARI1}. As particular members of the broader family of localized waves \cite{SAARI1,SAARI2,BOOK1,BOOK2}, their spatiotemporal characteristics and experimental generation have deserved decades of research. Most of this research is focused on localized waves without orbital angular momentum (OAM) \cite{SAARI1,SAARI2,BOOK1,BOOK2}, and more recently on synthesizing one-dimensional localized waves, or space-time light sheets \cite{FLORIDA1,FLORIDA2}, which cannot carry OAM. This fact could seem surprising since the interest in OAM of light started and developed over the same decades, at first for monochromatic light\cite{ALLEN,YAO}, and later for ultrashort pulses \cite{SHVEDOV,YAMANE,PORRAS1,PORRAS2,PORRAS3}. Only recently vortex-carrying X-waves and other diffraction-free pulses with OAM have started to gain attention \cite{CONTI1,CONTI2,PORRAS4,ORNI,PANG}.

Localized waves with cylindrical symmetry are built as coherent superpositions of diffraction-free Bessel beams of order $m=0$ (without OAM) or $m\neq 0$ (with OAM) of different frequencies $\omega$ and weights $f(\omega)$. The integer $m$ is the topological charge of the vortex in their center. Localized waves are dispersion-free because the $\omega$-dependent cone angle of the Bessel beams, $\theta(\omega)$, is such that the axial wave number (projection of the wave vector onto the propagation direction, say $z$) follows the linear variation law with frequency $k_z(\omega)=(\omega/c)\cos{\theta(\omega)} =a +\omega /v_g$, where $a$ is an arbitrary constant, and $v_g$ the group velocity. X-waves are characterized by $v_g>c$ and $a=0$, and generic superluminal localized waves \cite{SAARI2}, called here generic X-waves (GX-waves for short) are characterized by $v_g>c$ and $a\neq 0$.

The recent theoretical studies on \cite{CONTI1,CONTI2,PORRAS4} on X-waves with OAM have unveiled a rather complex OAM-temporal coupled structure obeying certain universal constraints. In \cite{CONTI1,CONTI2} the OAM-temporal couplings in the immediate vicinity of the vortex singularity are examined. With the same spectrum $f(\omega)$, the number of oscillations and their frequency increase with the magnitude of the topological charge, $|m|$. On the other hand, the bright ring surrounding the vortex is studied in \cite{PORRAS4}. In X-waves built with the same $f(\omega)$, X-waves have increasing duration at this ring when $|m|$ is increased, while the local frequency of the oscillations does not experience appreciable change with $|m|$, but is solely determined by the particular spectrum $f(\omega)$ of Bessel beams, e. g., the central frequency $\omega_f$ for bell-shaped spectra. The dependence of the duration of X-waves on $m$ originates from the lower bound $\Delta t \gtrsim |m|/\omega_f$ satisfied by all X-waves at their bright ring.

\begin{figure}[!b]
\begin{center}
\includegraphics*[height=4.4cm]{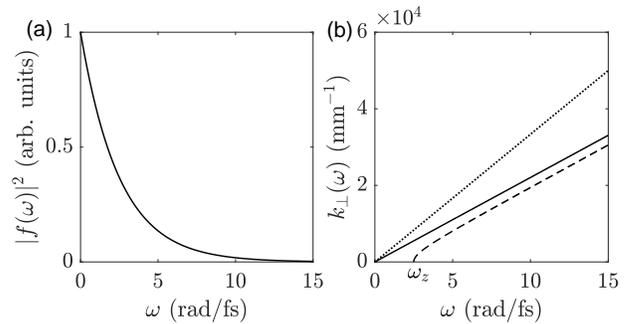}
\end{center}
\caption{\label{Fig1} (a) Spectral density of the broadband Bessel beam spectrum $f(\omega)=\exp(-\epsilon \omega)$ with $\epsilon=0.2$ fs, spanning the visible spectrum and beyond. (b) Transversal wave numbers of the X-wave with $v_g=0.0004$ mm/fs (solid line), of the GX-wave with of the same velocity and $\omega_z=2.5$ rad/fs ($a=2.083\times 10^3$ mm$^{-1}$) (dashed curve), and $\omega/c$ (dotted line).}
\end{figure}

Thus, to date, there is no a complete picture of the spatiotemporal structure of OAM-carrying X-waves except in the vicinity of the vortex and the bright ring. Here we describe the spatiotemporal structure of X-waves and GX-waves with OAM and with the broadband spectrum $f(\omega)=e^{-\epsilon \omega}$, where $\epsilon$ is a small quantity for $f(\omega)$ to cover from a dc component to the optical spectrum and beyond, as in Fig. \ref{Fig1}(a). Without OAM, the fundamental X-wave is non-oscillatory in time, and therefore is of little interest in optics; indeed only X-waves with narrower spectra about microwave \cite{MUGNAI} and optical frequencies \cite{SAARI1}, called Bessel-X waves, \cite{SONAJALG} have been generated. However, the introduction of OAM eliminates any component about $\omega=0$ and induces what can be qualified as intrinsic temporal oscillations associated with OAM, whose frequency can be tuned in any range of the electromagnetic spectrum. The previously described OAM-temporal couplings \cite{CONTI1,CONTI2,PORRAS4} arise here naturally as a result of the whole spatiotemporal coupled structure of these X-waves with OAM.

The OAM-induced oscillations feature a number of zeros approximately equal to $|m|$, and hence $|m|/2$ oscillations. Since the same number of oscillations fill the inside of the X arms at any radial distance from the vortex singularity, their frequency is red shifted radially outwards down to zero at infinity (with zero amplitude). With increasing $|m|$, the entire X-wave is blue shifted, at the same time that the bright ring is displaced outwards, resulting in a frequency at this ring independent of the topological charge. The structure of GX-waves closely resembles that of X-waves in their inner part, but the number of oscillations becomes increasingly larger than those of X-waves of same OAM towards the periphery. In addition, the oscillations spread out of temporally out of the X arms in the outer radial part, and their frequency approaches a non-zero constant value.

The duration of the broadband X-waves at their bright ring is found to coincide with the lower bound $|m|/\omega_f$ described in \cite{PORRAS4}. Therefore broadband X-waves are minimal X-waves capable to carry a given amount of OAM. Being diffraction-free, dispersion-free, and minimal OAM carriers (and other properties such as self-healing behavior and turbulence resistance \cite{LI}), broadband X-waves appear as optimum waves modes in many applications, particularly in superdense, multichannel free-space communications \cite{HUANG}, and free-space quantum communications systems \cite{PATERSON,ZHANG}, commonly based on Laguerre-Gauss-type modes.

\begin{figure*}[!]
\begin{center}
\includegraphics*[height=4.8cm]{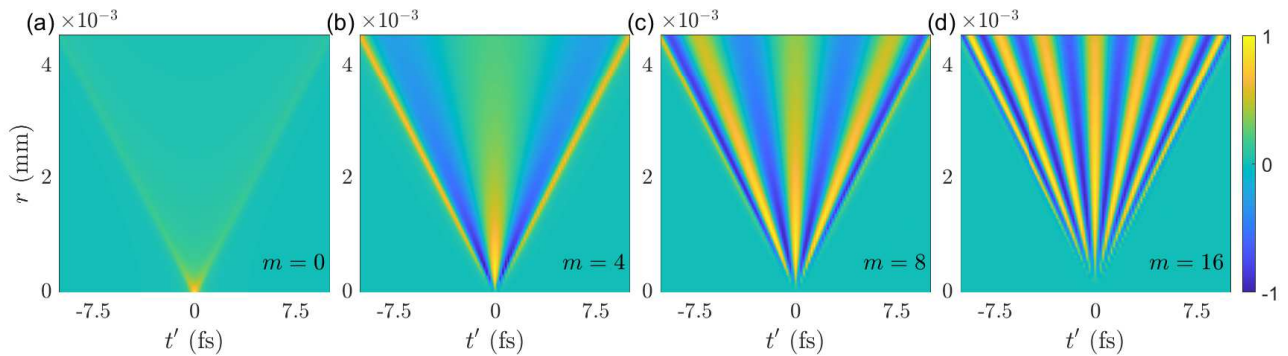}
\end{center}
\caption{\label{Fig2} Real electric field of X-waves with $v_g=0.0004$ mm/fs, $\epsilon =0.2$ fs and the indicated topological charges. All fields are normalized to their respective peak values.}
\end{figure*}

\section{Superluminal localized waves with OAM}

A quite general expression of cylindrically symmetric localized waves with OAM is $E(r,\varphi,t,z)=E(r,t,z)e^{im\varphi}$, with
\begin{equation}\label{LW}
E(r,z,t) = \frac{e^{iaz}}{\pi}\int_{0,\,k_\perp\mbox{\small real}}^\infty \!\!\!d\omega f(\omega) J_m[k_\perp(\omega) r] e^{-i\omega t'}\, ,
\end{equation}
where $(r,\varphi,z)$ are cylindrical coordinates, $t'=t-z/v_g$ is the local time for the group velocity $v_g$, and $k_\perp(\omega)= (\omega/c)\sin\theta(\omega) = \sqrt{(\omega/c)^2 - k_z^2(\omega)}$ is the transversal wave number (modulus of the transversal projection of the wave vector), or, given the linear variation of $k_z(\omega)$,
\begin{equation}\label{KPERP}
k_\perp(\omega) = \sqrt{\left(\frac{\omega}{c}\right)^2 - \left(a+\frac{\omega}{v_g}\right)^2}\,.
\end{equation}
The integral has been limited to nonnegative positive frequencies to yield the analytical signal complex representation of the electric field, whose real part is the real field. The limitation to real $k_\perp(\omega)$ expresses the restriction that the axial wave number $a+\omega/v_g$ cannot be higher than $\omega/c$. The electric field of X-waves ($a=0$, $v_g>c$) does not depend on $z$. The transversal wave number is the straight line $k_\perp(\omega)=(\sin\theta/c)\omega$ crossing the origin $\omega=0$ of slope $\sin\theta/c$ [Fig.\ref{Fig1}(b)], corresponding to a constant cone angle $\theta=\sin^{-1}\left(c \sqrt{1/c^2-1/v_g^2}\right)$. For GX-waves ($a \neq 0$, $v_g>c$) the electric field oscillates with axial period  $2\pi/a$. The transversal wave number $k_\perp(\omega)$ is a branch of hyperbola starting at some positive frequency $\omega_z$ (the other branch is entirely in $\omega<0$), the same asymptotic slope $\sin\theta/c$ as the X-wave of the same $v_g$ [Fig. 1(b)], and a $\omega$-dependent cone angle approaching $\theta$ at large $\omega$. We will fix the frequency $\omega_z$ as an important frequency of GX-waves by setting
\begin{equation}
a=\omega_z\left(\frac{1}{c}-\frac{1}{v_g}\right)\,,
\end{equation}
so that $\omega_z=0$ specifies an X-wave, and $\omega_z>0$ a GX-wave, see Fig \ref{Fig1}(b).

\subsection{Broadband X-waves with OAM}

Taking, as in \cite{LU}, the broadband exponential spectrum $f(\omega)=\exp(-\epsilon \omega)$, and setting $a=0$ ($\omega_z=0$) and $k_\perp(\omega) =(\sin\theta/c)\omega$, the integral in (\ref{LW}) can be carried out to yield
\begin{eqnarray}\label{XW}
E(r,z,t) &=&\frac{1}{\pi \sqrt{(\epsilon+it')^2 + \left(\frac{\sin\theta}{c}{r}\right)^2}} \times \nonumber\\
         &\times&  \frac{\left(\frac{\sin\theta}{c}{r}\right)^{|m|}}{\left[\sqrt{(\epsilon+it')^2 + \left(\frac{\sin\theta}{c}{r}\right)^2}+\epsilon +i t'\right]^{|m|}}
\end{eqnarray}
(and multiplied by $(-1)^{|m|}$ if $m<0$). Equation (\ref{XW}) was already derived in the pioneering work of Lu and Greenleaf \cite{LU}: The integrals involved in obtaining the expression for the fundamental, OAM-free X-wave were performed with Bessel functions of arbitrary order $m$. The expression with $m\neq 0$, however, received no attention either in that work or subsequently, to the best of our knowledge. 

The electric field approaches zero as $r^{|m|}$ close to the vortex singularity, and as $1/r$ at large enough distances, carrying then infinite energy, as the OAM-free X-wave. In Figs. \ref{Fig2}(a-d) the real electric field of X-waves without and with OAM can be compared. There is no light at times immediately outside the X arms, so that the duration at each radial distance is the time between the X arms, $2\Delta t= 2(\sin\theta/c)r$.

\begin{figure}[!]
\begin{center}
\includegraphics*[height=11.3cm]{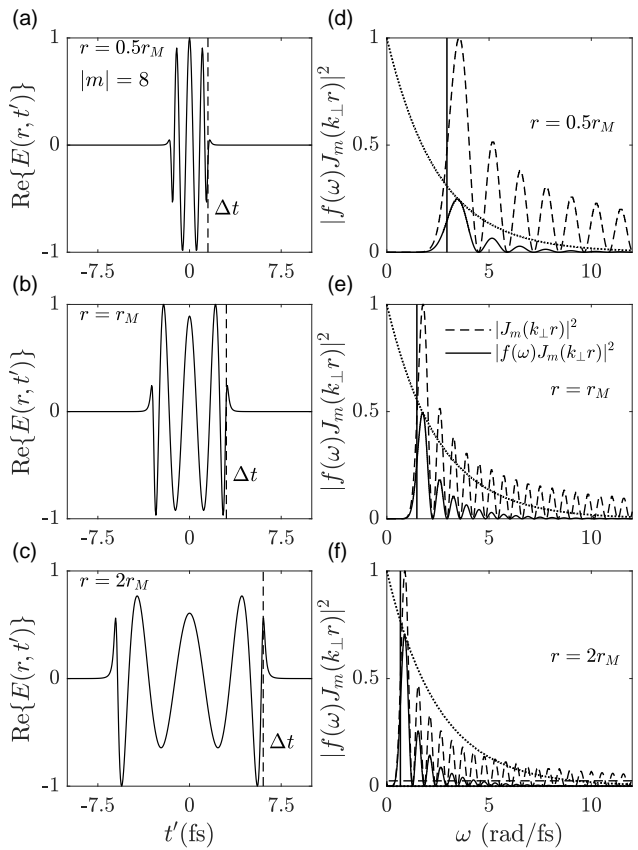}
\end{center}
\caption{\label{Fig3} (a-c) Real electric field of X-waves with $v_g=0.0004$ mm/fs, $\epsilon =0.2$ fs and $|m|=8$ at (a) $r=0.5 r_M$, (b) $r=r_M$ and (c) $r=2r_M$, where $r_M$ is the radial distance of maximum fluence. $\Delta t =(\sin\theta/c) r$ indicates the location of the X arms, where the X-wave terminates. All fields are normalized to their peak values. (d-f) Respective power spectra (solid curves), the Bessel factor $J_m(k_\perp r)$ (dashed curves), and the broadband spectrum factor $f(\omega)= e^{-\epsilon \omega}$ (dotted curves). The vertical lines indicate the central frequency about $t'=0$, $\omega_c(r)$, coinciding with the first rise of the Bessel functions in each case.}
\end{figure}

The most evident and relevant difference is that with $m=0$ the electric field is a unipolar pulse that splits into two unipolar pulses, while with $m\neq 0$ the temporal pulse shape at any radial distance has approximately $|m|$ zeros, or more precisely, the smallest even number greater or equal to $|m|$, e.g., 2 for $|m|=1,2$; 4 for $|m|=3, 4$, and so on. The number of oscillations may slightly differ depending on the particular criterion.  According to the instantaneous frequency analysis below, the number of oscillations is $|m|/2$ irrespective of whether $m$ is even or odd. These oscillations are more clearly seen in Figs. \ref{Fig3}(a-c) for $|m|=8$ at different radial distances. Thus, the increase of the number of oscillations with the magnitude of the topological charge does not only pertain to the vicinity of the vortex, as reported in \cite{CONTI1}, but to the whole X-wave. These are intrinsic oscillations associated with OAM, and result from the fact that the inverse Fourier transform $\int_0^\infty J_m\left[(\sin\theta/c)\omega r \right]e^{-i\omega t}d\omega$ has these zeros and oscillations. The exponential $e^{-\epsilon \omega}$ does not remove them, but only makes those at the trailing and leading parts of the pulse to have smaller and smaller amplitude towards the vortex center, as seen in the temporal shapes Figs. \ref{Fig3} from (a) to (c). This softening is the result of the increasing apodization of the Bessel function by the exponential spectrum towards the vortex center, as observed in the respective spectral densities in Figs. \ref{Fig3} from (d) to (f).

With a fixed number of oscillations at all radii within a linearly increasing time interval $2\Delta t = 2(\sin\theta/c) r$ between the X arms, their frequency must decrease approximately inversely proportional to $r$. Also, with a number of oscillations proportional to $|m|$, their frequency at any particular radius $r$ must increase proportionally to $|m|$.

\begin{figure}[!]
\begin{center}
\includegraphics*[height=4.1cm]{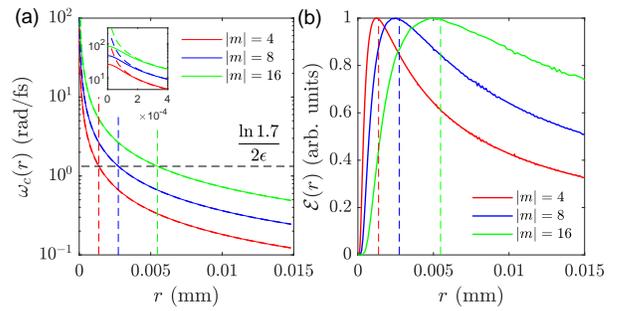}
\end{center}
\caption{\label{Fig4} (a) Central instantaneous frequency of the oscillations, $\omega_c(r)$, as a function of the radius $r$ for the indicated values of $|m|$ of X-waves with $v_g=0.0004$ mm/fs and $\epsilon =0.2$ fs, as given by the exact expression (\ref{WC}) (solid curves) and the approximate expression (\ref{WC2}) (dashed curves). They are almost indistinguishable except in the immediate vicinity of the vortex singularity. This region is enlarged in the inset. (b) Radial profiles of fluence of the same X-waves, numerically evaluated from (\ref{F}), and normalized to their peak values. In (a) and (b) the vertical lines are $r_M$ given by (\ref{RM}), locating approximately the radii of maximum fluence. The horizontal line in (a) helps to visualize that the central instantaneous frequency $\omega_f=\ln(1.7)/2\epsilon$ is the same at the respective radii of maximum fluence, i. e., independent of $m$.}
\end{figure}

For a more quantitative analysis, we consider the instantaneous frequency, defined as $\omega_c =- d\,\mbox{arg} E/dt'$, which can also be evaluated from
\begin{equation}
\omega_c(r,t') = -{\rm Re}\left\{\frac{\int_0^\infty e^{-\omega(\epsilon+it')}J_m [k_\perp(\omega) r]\omega d\omega}{\int_0^\infty e^{-\omega(\epsilon+it')}J_m [k_\perp(\omega) r] d\omega}\right\}\,,
\end{equation}
yielding
\begin{equation}\label{WCE}
\omega_i(r,t') = {\rm Re}\left\{\frac{|m|\sqrt{(\epsilon+it')^2 + \left(\frac{\sin\theta}{c}r\right)^2}+(\epsilon+it')}{(\epsilon+it')^2+\left(\frac{\sin\theta}{c}r\right)^2}\right\}\,.
\end{equation}
Simple inspection shows that $\omega_i(r,t')$ takes a minimum value 
\begin{equation}\label{WC}
\omega_c(r)\equiv \omega_i(r,t'=0) = \frac{|m|\sqrt{\epsilon^2 + \left(\frac{\sin\theta}{c}r\right)^2}+\epsilon}{\epsilon^2+\left(\frac{\sin\theta}{c}r\right)^2}\,,
\end{equation}
at $t'=0$, or central instantaneous frequency, that remains almost constant in time except in the vicinity of the X arms, and is
plotted as a function of $r$ for several values of $|m|$ in Fig. \ref{Fig4}(a). As the minimum frequency at each radial distance, it coincides with the first rise of the Bessel function in the spectrum, indicated by vertical lines in Figs. \ref{Fig3}(d-f). In fact a good approximation to (\ref{WC}) can be derived by equating the argument $x=k_\perp r= (\sin\theta/c)\omega r$ of the Bessel function $J_m(x)$ to that of the first rise of the Bessel function, $x \simeq |m|$, which yields
\begin{equation}\label{WC2}
\omega_c(r)\simeq \frac{|m|}{(\sin\theta/c) r}\,.
\end{equation}
This approximate equality is seen in Fig. \ref{Fig4}(a) to fit accurately (\ref{WC}) except in a tiny radial region [see inset in Fig. \ref{Fig4}(a))] about the vortex (compared to the radius of maximum X-wave energy, $r_M$). Thus, except in that region, the frequency $\omega_c(r)$ is independent of the particular broadband spectrum defined by $\epsilon$, and is inversely proportional to $r$, as expected. For $r\rightarrow 0$, the exact formula (\ref{WC}) yields the finite value
\begin{equation}\label{WC0}
\omega_c (0) = \frac{|m|+1}{\epsilon}\,,
\end{equation}
which is similar to the result in Ref. \cite{CONTI1,CONTI2}. From the latter relation and (\ref{WC2}), we also conclude that the blue shift proportional to the magnitude of the topological charge does not only takes place in the vicinity of the vortex, but affects the whole X-wave.

Equation (\ref{WCE}) for the instantaneous frequency at any time allows us to evaluate the number of oscillations within the X arms. In the same way as $\omega_c(r)$, $\omega_i(r,t')$ in (\ref{WCE}) turns out to be almost independent of $\epsilon$, except in the vicinity of the X-arms, and to be approximately given by
\begin{equation}
\omega_i(r,t') \simeq \frac{|m|}{\sqrt{\left(\frac{\sin\theta}{c}r\right)^2- t^{\prime 2}}}\,
\end{equation}
provided that $|t'|<\Delta t = (\sin\theta/c)r$, i. e., within the X arms. Averaging between in this time interval (integrating and dividing by $2\Delta t$) yields an average instantaneous frequency at each radius as
\begin{equation}
\bar \omega_i(r) = \frac{\pi}{2}\frac{|m|}{(\sin\theta/c)r}\,.
\end{equation}
One can then evaluate the number of oscillations as the full duration $2\Delta t$ over the average period $2\pi/\bar\omega_i(r)$, resulting in a number of oscillations equal to $|m|/2$.

The global blue shift of X-waves with $|m|$ might lead one to think that the whole X-wave becomes bluer and bluer with increasing magnitude of the topological charge. However this is only the case in the immediate vicinity of the vortex, as seen in (\ref{WC0}). According to (\ref{WC2}), valid out of this region, any given frequency is displaced radially outwards as $|m|$ is increased. As shown below, the radius of maximum energy density (fluence), or bright ring to a time-integrating detector, is also displaced with increasing $|m|$ in such a way that the frequency of the bright ring is independent of $m$, and in this sense it can be said that X-waves are of the same color irrespective of their OAM.

The fluence is given by ${\cal E}(r)=\int_{-\infty}^\infty (\mbox{Re}E)^2dt' =\frac{1}{2}\int_{-\infty}^\infty|E|^2 dt'$, or, in terms of the spectral density, by
\begin{equation}\label{F}
{\cal E}(r)=\frac{1}{\pi}\int_0^\infty e^{-2\epsilon\omega}|J_m(k_\perp(\omega) r)|^2 d\omega \,,
\end{equation}
which is plotted in Fig. \ref{Fig4}(b) for several values of $|m|$ for illustration. Although the above integral does not admit analytical integration, detailed numerical inspection shows that the area of the product $e^{-2\epsilon\omega}$ and $|J_m(k_\perp r)|^2$ is maximum, and then the fluence, at the radius $r_M$ where the frequency of the first rise of $|J_m(k_\perp r)|^2$, i. e. $\omega_c(r_M)$, coincides with the frequency $\omega_f$ at which the broadband spectral density $e^{-2\epsilon\omega}$ has decayed by about 1/1.7 =0.588 its value at $\omega=0$, i. e., $\omega_f \simeq \ln (1.7)/2\epsilon$. With this relative position, the Bessel function is not too damped and not too oscillatory, as exemplified in Figs. \ref{Fig3} from (d) to (f). The frequency at the bright ring, $\omega(r_M)\simeq \omega_f$, is then independent of $m$, as illustrated in Fig. \ref{Fig4}(a).
The fact that the frequency at the bright ring is solely determined by $f(\omega)$ was also reported in \cite{PORRAS1} for X-waves with bell-shaped spectra, and for ultrashort Laguerre-Gauss pulses in \cite{PORRAS1,PORRAS2,PORRAS3}. From (\ref{WC2}) equated to $\omega_f$, we obtain the radius of the bright ring as
\begin{equation}\label{RM}
r_M\simeq \frac{|m|}{\omega_f (\sin \theta/c)}\,,
\end{equation}
which is proportional to $|m|$, and provides a good approximation to the exact radius, see Fig. \ref{Fig4}(b).

Also in Ref. \cite{PORRAS1}, the lower bound to the duration at the bright ring of X-waves carrying $m$ units of OAM is established as $\Delta t \gtrsim |m|/\omega_f$ (half duration). Broadband X-waves have just the minimum duration $\Delta t=(\sin\theta/c) r_M = |m|/\omega_f$, and are therefore the minimal X-waves capable to carry $m$ units of OAM.

\subsection{Broadband GX-waves with OAM}

\begin{figure}[!]
\begin{center}
\includegraphics[height=4.6cm]{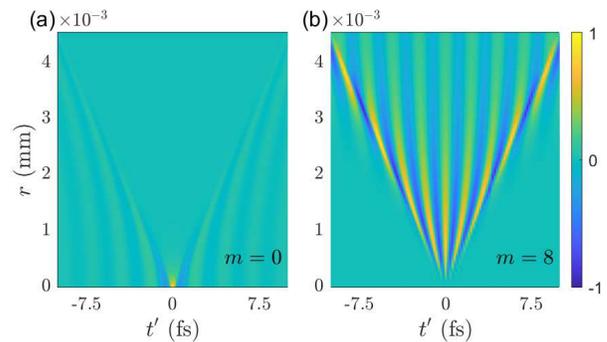}
\end{center}
\caption{\label{Fig5} Real electric field of GX-waves with $v_g=0.0004$ mm/fs, $\omega_z=2.5$ rad/fs, and $\epsilon =0.2$ fs (a) without OAM and (b) with OAM.}
\end{figure}

Integral (\ref{LW}) with the exponentially decaying Bessel beam spectrum and $k_\perp(\omega)$ in (\ref{KPERP}) with $\omega_z\neq 0$ ($a\neq 0$) cannot be performed analytically, but the spatiotemporal structure of these GX-waves can be easily understood from that of the X-wave with $a=0$ of the same group velocity and vorticity. Without OAM, GX-waves and X-waves differ substantially [compare Fig. \ref{Fig2}(a) with Fig. \ref{Fig5}(a)] because the spectrum $f(\omega)J_m(k_\perp r)$ of GX-waves with $m=0$ is highly peaked at the positive cut-off frequency $\omega_z$ (where $k_\perp r=0$) responsible for the infinite temporal oscillations observed at any radial distance. For GX-waves with OAM, however, the spectrum $f(\omega)J_m(k_\perp r)$ vanishes at $\omega_z$, which removes these oscillations, and makes the GX-wave with OAM to resemble much more the X-wave with OAM [compare Fig. \ref{Fig2}(c) with Fig. \ref{Fig5}(b)].

\begin{figure}[!]
\begin{center}
\includegraphics*[height=10.7cm]{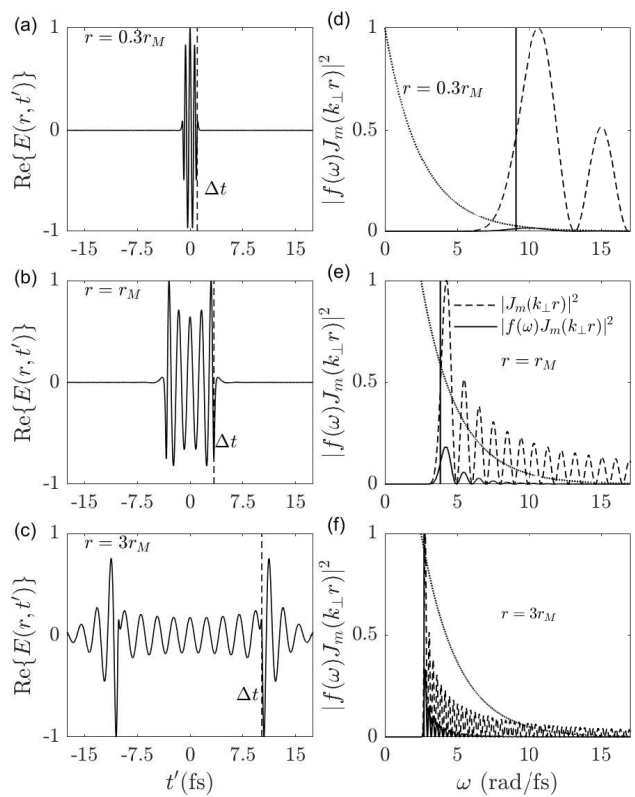}
\end{center}
\caption{\label{Fig6} (a-c) Real electric field of GX-waves with $v_g=0.0004$ mm/fs, $\omega_z=2.5$ rad/fs, $\epsilon =0.2$ fs and $|m|=8$ at $r=0.3 r_M$, $r=r_M$ and $r=3r_M$, where the radius of maximum fluence is given by (\ref{RM2}). $\Delta t =(\sin\theta/c) r$ indicates the location of the X arms. (d-f) Respective power spectra (solid curves), the Bessel factor $J_m(k_\perp r)$ (dashed curves), and the broadband spectrum factor $f(\omega)= e^{-\epsilon \omega}$ (dotted curves). The vertical lines indicate the central frequency $\omega_c(r)$ given by (\ref{WC3}), and coinciding with the rise of the Bessel functions in each case.}
\end{figure}

In the vicinity of the vortex the number of oscillations is indeed the same as that of the X-wave of the same vorticity [Fig. \ref{Fig6}(a)]. This feature can be understood from the fact that the slope of $k_\perp(\omega)$ at high frequencies is the same as for the X-wave [Fig. \ref{Fig1}(b)], and that these high frequencies are located in the vicinity of the vortex. The cut-off frequency $\omega_z$ plays negligible role in the spectrum at these distances [Fig. \ref{Fig6}(d)]. Moving towards the periphery the number of oscillations becomes gradually larger than that of the X-wave, growing without bound and approaching the constant frequency $\omega_z$ [Fig. \ref{Fig6}(b)] because the spectrum approaches $\omega_z$ [Fig. \ref{Fig6}(e)] with increasing radius. At large enough radius [Fig. \ref{Fig6}(c)] the oscillations go beyond the X arms, as for the GX-wave without OAM, because the spectrum becomes dominated at these large distances by the cut-off frequency [Fig. \ref{Fig6}(f)].

\begin{figure}[b!]
\begin{center}
\includegraphics*[height=4.2cm]{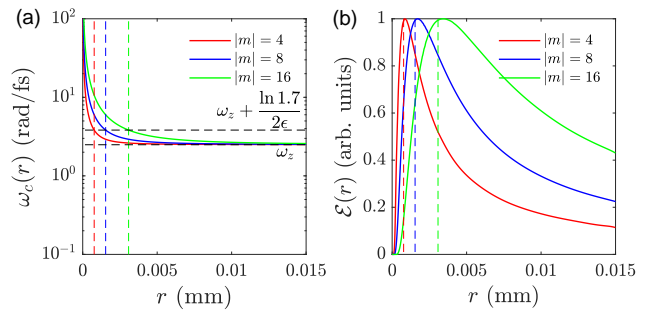}
\end{center}
\caption{\label{Fig7} (a) Central instantaneous frequency of the oscillations, $\omega_c(r)$, as a function of the radius $r$ for the indicated values of $|m|$ of GX-waves with $v_g=0.0004$ mm/fs, $\epsilon =0.2$ fs, and $\omega_z=2.5$ rad/fs as given by the approximate expression (\ref{WC3}). In the immediate vicinity of the vortex singularity $\omega_c(r)$ does not approach infinity but to $\omega_c(0)\simeq (m+1)/\epsilon$. (b) Radial profiles of fluence of the same GX-waves, numerically evaluated from (\ref{F}), and normalized to their peak values. In (a) and (b) the vertical lines are $r_M$ given by (\ref{RM}), locating approximately the radii of maximum fluence. The horizontal lines in (a) helps to visualize that the central instantaneous frequency $\omega'_f=\omega_z + \ln(1.7)/2\epsilon$ is the same at the respective radii of maximum fluence, i. e., independent of $m$, and that the frequency approaches $\omega_z$ at large radius.}
\end{figure}

An approximate expression for the radial distribution of frequencies at the GX-wave temporal center, $t'=0$, can be obtained as for X-waves. Equating the argument of $J_m(x)$, with $x=k_\perp(\omega)r$ and $k_\perp(\omega)$ given by (\ref{KPERP}), to the location of the first rise of $J_m(x)$, $x\simeq |m|$, we obtain a quadratic equation in $\omega$ whose positive solution is
\begin{equation}\label{WC3}
\omega_c(r) \simeq \omega_z + \frac{c^2}{\sin^2\theta} \left( -\frac{a}{c} +\sqrt{\frac{a^2}{c^2}+ \frac{\sin^2\theta}{c^2}\frac{m^2}{r^2}} \right),
\end{equation}
which is independent of $\epsilon$, as for X-waves, and approaches $\omega_z$ at large radius, as expected [Fig. \ref{Fig7}(a)]. Expression (\ref{WC3}) only fails in the close vicinity of the vortex, where the frequency becomes $\epsilon$-dependent and reaches approximately the same value $\omega_c(0)\simeq (|m|+1)/\epsilon$ as for X-waves.

We also observe that the central frequency $\omega_c(r_M)$ at the ring of maximum fluence is substantially independent of $|m|$ [Fig. \ref{Fig7}(a)] and determined solely by the frequency where $|f(\omega)|^2= e^{-2\epsilon \omega}$ has decayed approximately the same value $1/1.7=0.588$ as for X waves from its value at the cut-off frequency $\omega_z$, i. e., $\omega'_f =\omega_z + \ln(1.7)/2\epsilon$. Equating $\omega'_f$ to $\omega_c(r)$ in (\ref{WC3}), we obtain, after some algebra, the radius of maximum fluence as
\begin{equation}\label{RM2}
r_M\simeq \frac{|m|}{\sqrt{\omega_f^2\frac{\sin^2\theta}{c^2} + 2\omega_f\frac{a}{c}}}\,.
\end{equation}
Expression (\ref{RM2}) provides a reasonably good approximation to the radius of maximum fluence [Fig. \ref{Fig7}(b)]. This radius continues to be proportional to $|m|$, but is smaller than for X-waves.

\section{Conclusions}

We have described the strongly coupled spatiotemporal structure of broadband superluminal localized waves with OAM. Temporal oscillations at all the frequencies in the broadband spectrum are displayed at different radii between the X arms, with a fixed number of oscillations dictated by OAM. A steep red shift with radial distance in conjunction with a pronounced blue shift of the whole X-wave or GX-wave with the magnitude of the topological charge results in an invariant color at the ring of maximum energy density, whose frequency is only determined by the spectrum of Bessel beams (which would directly be related to the spectrum of the laser source).  

The practical generation of these diffraction-free, dispersion-free, self-healing, and minimal OAM-carrier wave modes would have an impact in  obvious applications such as free-space, classical and quantum communication systems \cite{HUANG,PATERSON,ZHANG} currently using Laguerre-Gauss modes as OAM carriers. In strong-field, nonperturbative light-matter interactions such as in the generation of high harmonics and attosecond pulses with OAM, \cite{HERNANDEZ,GARIEPY,REGO} replacing the standard Laguerre-Gauss modes with OAM-carrying X-waves would enormously lengthen the depth of focus where light and matter can interact with a more uniform axial field.

We have deliberately left subluminal localized waves with OAM aside since they are expected to follow completely different rules as the transversal wave number dispersion is not hyperbolic but elliptical. This analysis is deferred for further work.

M.A.P. acknowledges support from Projects of the Spanish Ministerio de Econom\'{\i}a y Competitividad No. MTM2015-63914-P, and No. FIS2017-87360-P.

\end{document}